\documentclass[MSNbibl,nameyear,dvips]{arxstspdf}
\usepackage{flushend}
\usepackage{stfloats}

\volume{26}
\issue{4}
\pubyear{2011}
\firstpage{471}
\lastpage{478}
\doi{10.1214/11-STS379}

\begin{document}
\begin{frontmatter}

\title{Election Forensics and the 2004 Venezuelan Presidential Recall
Referendum as a Case Study}
\runtitle{Election Forensics}

\begin{aug}
\author[a]{\fnms{Alicia L.} \snm{Carriquiry}\corref{}\ead[label=e1]{alicia@iastate.edu}}
\runauthor{A. L. Carriquiry}

\affiliation{Department of Statistics}

\address[a]{Alicia L. Carriquiry is Distinguished Professor, Department
of Statistics, Iowa State University, Ames, IA 50011-1210, USA \printead{e1}.}

\end{aug}

%
\begin{abstract}
A referendum to recall President Hugo Ch\'avez was held in Venezuela in
August of 2004. In the referendum, voters were
to vote YES if they wished to recall the President and NO if they
wanted him to continue in office.
The official results were 59\% NO and 41\% YES. Even though the
election was monitored by various
international groups including the Organization of American States and
the Carter Center
(both of which declared that the referendum had been conducted in a
free and transparent manner),
the outcome of the election was questioned by other groups both inside
and outside of Venezuela.
The collection of manuscripts that comprise this issue of {\it
Statistical Science} discusses the
general topic of election forensics but also focuses on different
statistical approaches to explore, post-election,
whether irregularities in the voting, vote transmission or vote
counting processes could be detected in the 2004 presidential recall
referendum. In this introduction to the Venezuela issue, we discuss the
more recent literature on post-election auditing, describe the
institutional context for the 2004 Venezuelan referendum, and briefly
introduce each of the five contributions.

\end{abstract}

%

\vspace*{-3pt}
\end{frontmatter}

\section{Introduction} \label{sec:intro}

In every democracy, citizens have the opportunity to participate in
elections at different levels. It is critically important that
elections be free and fair, by which we mean that institutions and
safeguards to guarantee full access of eligible voters and of
candidates must be in place. In most advanced democracies we can
reasonably assume that this will be the case (except in localized
instances of malfeasance or non-intentional error) but with the advent
of many new democracies around the world, opportunities for
irregularities of all kinds have multiplied (Mebane, \citeyear{2007Mebane}). 

Since approximately the 1960s (Hyde, \citeyear{2011Hyde}), 
the practice of inviting international observers to monitor the
electoral process as it takes place has become almost standard
practice. International organizations such as the United Nations or the
Organization of American States, alongside non-governmen\-tal groups such
as the Carter Center, have often been asked to produce a ``seal of
approval'' for elections that were expected to be contested. The 2004
Venezuelan Referendum to Recall the President was one such election as
we discuss below. While ensuring that elections are free and fair has
been the main focus of these organizations in recent years, less has
been done about the question of whether the election outcome reflects
the intentions of the electorate.

Election {\it monitoring}, typically conducted by observation of the
electoral process as it occurs in the field, is aimed at detecting
irregularities such as voter access and the integrity of ballot boxes,
but does not address the issue of election accuracy. We say that an
election is accurate when the outcome of the election is consistent
with the preferences of voters. By {\it accurate} we do not mean {\it
perfect}; in every election, there will be some differences between the
official vote count and the actual votes that were cast by the
electorate. These differences are often inconsequential, in that they
do not affect the outcome of the election. When the differences are
large enough to determine an outcome that is not reflective of voters'
intentions, then the election is said to be inaccurate (Mebane, \citeyear{2007Mebane}).

The consequences of conducting elections that are provably flawed,
either by mistake or by malfeasance, can be costly from a political,
social and even economic point of view. A winner who is perceived to be
``illegitimate'' might not be able to gain the respect of disgruntled
voters and in extreme cases, might be unable to lead (as was the case
with the presidential elections of 2001 in Bangladesh, European Union,
\citeyear{2001European}). 
Further, after an election has been conducted, it is a challenge to
decide what to do even when significant irregularities have been
detected. An example close to home was the voting in Jasper County,
South Carolina, during the presidential election of 2000. While it was
clear that tampering with voting machines had occurred, the legal
battles that ensued post-election did not lead to widely accepted
rulings (Jacobson and Rosenfeld, \citeyear{2002Jacobson}). 
(The discussions came to an end once it was argued that even if the
fraudulent voting in Jasper County had not occurred, South Carolina
would have been a Bush state anyway.) In practical terms, any remedial
measure implemented after an election has been declared inaccurate has
its own limitations and thus, fixing an election problem after the fact
is typically very difficult.

The introduction of voting machines of different kinds has created some
uncertainty in the outcome of elections, even in well-established
democracies such as the United States (Lehoucq, \citeyear{2003Lehoucq}).
Voting machines do not always produce printed reports, and when they do,
the reports are unsuitable for detecting voting irregularities (Dopp,
\citeyear{2009Dopp}). 
Data reports are rarely\vadjust{\goodbreak} produced in formats that enable statistical
analyses and often, the data are stored in proprietary file formats
that limit their usefulness. Serious errors in vote counting have been
documented; an example of a software failure that resulted in
inexplicably lost votes occurred during the 2008 presidential election
in several counties in California, where optical-count scanners
manufactured by Premier Election Solutions (formerly Diebold Election
Solutions) not only did not count votes but also deleted any signs that
the votes had been cast at all (Zetter, \citeyear{2008Zetter}). 
Voter-verifiable paper ballot records used by some electronic voting
machines are not error- and tamper-proof either (Balzarotti et al.,
\citeyear{2008Balzarotti}) 
because machines can be programmed to produce apparently matching
counts and paper reports that can be difficult to identify as fraudulent.

Over the last decade or two, there has been renewed interest in the
development of statistical me\-thodology that can be used in the course
of election audits (both pre- and post-election) to detect
irregularities and guarantee the integrity of the elections. The
ultimate goal of election auditing is to determine whether the winner
of the election has been called correctly. In the United States,
concerns about the legitimacy of elections reached a new height after
the 2000 presidential election, where a combination of flawed
administrative practices, voter suppression and other irregularities
threw into question the outcome in Florida. The idea that the results
of an election should be confirmed via some kind of post-election
manual tallying is becoming more accepted and has been
institutionalized in many states in the U.S. In California, for
example, the law now requires that ballots cast in no fewer than one
percent of the precincts in any election be manually recounted
(Saltman, \citeyear{2006Saltman}; Stark, \citeyear{2008Stark}). 
Many other state legislatures are considering bills that will also
require a post-election audit of anywhere between 1\% and 10\% of the
precincts, selected randomly in different ways. While these are
positive steps toward creating a system for carrying out post-election
audits in an ``objective'' way, no system that establishes a fixed
proportion of precincts to be audited can guarantee that a full manual
recount would confirm the outcome, with a sufficiently high
probability. Stark (\citeyear{2008Stark}, \citeyear{2010Stark}) 
proposes an approach to sample precincts that depends on the apparent
margin of victory, the number of precincts in the election, the number
of ballots cast in each precinct and the target level of confidence
that the real winner is called.\vadjust{\goodbreak} McCarthy et al. (\citeyear{2008Mccarthy}) 
propose a similar approach to selecting precincts for post-election
manual ballot re-count that depends on the power with which we wish to
identify the true election winner.
The collection of technologies and methods that can be used to assess
the legitimacy of elections is known as {\it election forensics}
(Mebane, \citeyear{2007Mebane}). 
In addition to the more statistical aspects of election forensics,
other published research has focused on the mechanics of post-election
auditing (Estok, Nevitte and Cowan, \citeyear{2002Estok}; Norden et al., \citeyear{2007Norden}). 

The manuscripts that are included in this issue of {\it Statistical
Science} propose different but complementary methods to collect,
analyze and interpret post-election data, and provide an overview of
the type of statistical tools that can be useful in evaluating the
integrity of an election. The motivation for all of the manuscripts
included in this set was the 2004 referendum carried out in Venezuela,
where voters were asked to vote YES or NO to the question of whether
President Hugo Ch\'avez should be recalled. Questions about the
election's accuracy were raised almost immediately and several groups
in Venezuela and abroad set out to analyze some of the data that became
available after the official results were announced. While the
manuscripts included in this issue suggest that various forms of
apparently intentional tampering seem to have occurred, other
contributions to the literature (Taylor, \citeyear{2005Taylor}; Weisbrot, Rosnick and Tucker, \citeyear{2004Weisbrot}) 
argue that the evidence is insufficient to conclude that the outcome of
the referendum was not correct.

The remainder of this introduction is organized as follows. First, we
briefly describe the 2004 presidential referendum in Venezuela. We then
discuss each of the five manuscripts that comprise the Venezuela
referendum set. We finish with a brief conclusions section.

\section{The 2004 Presidential Recall Referendum in Venezuela} \label{sec:rr}

In 1998, President Hugo Ch\'avez was elected President of Venezuela
with almost 58\% of the vote. As is required in Venezuela, the election
was organized by the CNE (Consejo Nacional Electoral), a body composed
of five individuals who must be confirmed by the legislative branch of
the Venezuelan government and that has the mission of ensuring that
elections are transparent and conducted according to the electoral
normatives. In 1999, a new national Constitution was enacted. The new
Constitution allowed for the conduct of presidential recall referenda
and established the protocol under which this type of referendum could
be conducted. In 2000, President Ch\'avez agreed to run for early
re-election and was re-elected to a new six-year term with almost 60\%
of the vote. While the integrity of the 2000 election already raised
some questions, no formal challenge was submitted. However, the
political situation in Venezuela continued to deteriorate and led to a
national strike that was resolved only when a~new CNE was established
in 2003 (with mediation by the Carter Center) and agreed to organize a
presidential recall referendum to be conducted in 2004.

The presidential recall referendum (RR) was conducted in August of
2004. This election was the first in which touch-screen voting machines
were ever used in a national election in Venezuela. A large proportion
(about 87\%) of all votes were cast in voting centers that used
touch-screen voting machines. The machines produced a vote confirmation
paper receipt for each voter, that were deposited in sealed ballot
boxes. Most of those paper voting records were not analyzed. The
machines were connected to the totalizing servers of the CNE via
telephone lines and transmitted the voting totals in each specific
machine to the servers. Two post-election audits of a subset of the
voting centers were conducted in cooperation with the Carter Center and
with the Organization of American States.

In the RR, participants could choose to vote SI (to recall the
president) or NO (to allow him to remain in his post). The official
count was 59\% for NO and 41\% for SI. The Carter Center declared that
the elections had been fair and transparent and their report pointed to
no major irregularities (Carter Center, \citeyear{2005aCarter}, \citeyear{2005bCarter}). 
Other non-governmental organizations, however, carried out analyses of
different sets of data arising from the election and the two
post-election audits and raised questions about the integrity of the RR.

\section{Analyses of the 2004 Presidential Recall Referendum} \label{sec:papers}

Here we briefly introduce the manuscripts that were accepted for
inclusion in the Venezuela issue. Four of the manuscripts, those by
Hausmann and Rigobon, Prado and Sans\'o, Pericchi and Torres, and Mart\'
in suggest that there is sufficient evidence to conclude that the
referendum was fraudulent. In contrast, the contribution by Jim\'enez,
while still declaring the RR outcome to be illegitimate, argues that
the available information is not conclusive enough to declare that the
official results are incorrect in the sense of Hausmann and Rigob\'on,
Prado and Sans\'o, or Pericchi and Torres.

\subsection{Delfino and Salas}

Delfino and Salas compare the proportion of YES votes in each voting
center with the proportion of voters registered in that center who had
signed the petition to request the referendum. The assumption
underlying their analysis is that most of the people who signed the
petition for the referendum are likely to vote YES. This is a plausible
assumption given that no signature collection centers were allowed
outside of Venezuela (yet voting centers were established in embassies
and consulates around the world) and that individuals participating in
the signature drive were easily identifiable by the government (but
votes cast during the referendum were secret).

Delfino and Salas find discrepancies between what one might expect
given the distribution of signing registered voters across centers and
the official proportion of YES votes in each of the centers. These
discrepancies appear to be larger in voting centers with touch-screen
voting machines than in non-com\-puterized centers. In the more populous
centers\break (where the number of registered voters is largest) the
relationship between the proportion of YES votes appears to be too
tightly associated with the proportion of registered voters who signed
the petition for the referendum.

Finally, the authors also compare the correlation between the
proportion of YES votes and the proportion of signatures in each voting
center, in two groups of centers defined by the proportion of
signatures, in different elections that were held in the same voting
centers between 1998 and 2004 and find inexplicable results. In all
other elections, the correlations are low in centers with small
percentages of registered voters who signed the petition, and higher in
centers with large proportions of signatures. In contrast, the
correlations were very high in both groups of voting centers during the
RR, a~result that cannot be easily explained.

\subsection{Pericchi and Torres}

Pericchi and Torres propose an approach to evaluate the integrity of
elections that relies on the New\-comb--Benford Law for first and second
digits and on a generalization\vadjust{\goodbreak} of the law for cases where the total
number of observations is capped. The main goal of the Pericchi and
Torres work is to develop a~new statistical tool that can be used in a
wide variety of applications to determine whether numerical outcomes
show irregularities.

The Newcomb--Benford Law establishes that the distribution of digits
(first, second, etc.) is not uniform. In principle, therefore, one can
compare the distribution of first significant digits, second
significant digits and so on to the distribution that would be expected
under the Newcomb--Benford Law, and use the discrepancy as a test
statistic for the hypothesis that the observed distribution of digits
is not due to tampering. The Newcomb--Benford Law holds only
asymptotically when the digits arise from aggregated unit-less counts
in small samples. Because the distribution of the first significant
digit depends on the size of the sample, Pericchi and Torres suggest
that the second Benford Law (or the law that refers to the distribution
of the second significant digit) has better statistical properties in
small samples.

The authors use several elections around the world to illustrate the
approach they propose for detecting departures from what would be
expected under no irregularities. They find that in all cases, the null
hypothesis that the frequency distribution of second significant digits
behaves according to Newcomb--Benford is not rejected. The only
exception is the 2004 Venezuelan presidential recall referendum, whe\-re
the Bayes factor for assessing the posterior probabilities of the null
model and the observed frequency distribution for the second
significant digit suggest that the Newcomb--Benford model is not
consistent with the observed frequencies.

A valuable contribution in this manuscript is the extension of the law
to cases where the total number of counts is bounded. Under a
restriction on the maximum number of counts, different voting precincts
(or other units) tend to have a constant number of voters. Pericchi and
Torres show that the frequency distribution of the second significant
digit is less sensitive to departures from the expected behavior under
the law even when the total number of units in each center is about the same.

\subsection{Prado and Sans\'o}

Prado and Sans\'o use exit poll data from two independent surveys
conducted during the RR by a~non-governmental organization called S\'
umate and by\break Primero Justicia, a political party in the opposition.
Both groups\vadjust{\goodbreak} collected voting information nationwide, by interviewing
voters as they exited the voting centers. To guarantee confidentiality,
respondents were asked to put their vote in a sealed envelope and in a
box similar to ballot boxes. No other information---gender, age,
socio-economic status or any other---was collected from the participants.

The forecasts that were obtained from both exit polls were in
remarkable agreement. Both predicted that the YES vote would win, with
about 60\% of the total votes cast. The sharp discrepancy with the
official CNE results, which reported that the NO vote was about 59\%,
was the motivation for the Prado and Sans\'o contribution.

Prado and Sans\'o carry out a simple analysis, that consists in
exploring whether the exit poll results are likely if we were to assume
that the official CNE election results are true. In other words, if in
fact the NO received almost 60\% of the vote, what is the probability
that we would observe the exit poll results that were observed in each
voting center? The calculation is tantamount to computing a $p$-value
for the hypothesis that the CNE results are correct, when using the
exit poll results as the test statistic. They find that for a large
proportion of the voting centers, these $p$-values are small, typically
below 0.02, providing some evidence against the assumption that the CNE
results are reliable. They note that this result is observed in centers
all over the country, whether large or small and for both computerized
and manual voting systems.

The disagreement between the official results and what the exit polls
predicted can be attributable to factors other than tampering by the
CNE. Prado and Sans\'o offer several alternative explanations for the
differences, but provide arguments that cast\break doubt on most of them.
Still, this type of analysis, while suggestive, in no way can lead to a
conclusion of tampering by the government, something with which the
authors readily agree. While inconclusive, the comparison of official
results and believable forecasts is a useful tool to at least call
attention to electoral events where irregularities may be present.

Some potentially significant drawbacks in the Pra\-do and Sans\'o
analyses are listed here. First, they have no information at all about
the proportion of voters who refused to respond to the exit poll
survey. Second, they do not know anything about the non-respondents. If
we are to reasonably assume that the probability of being a respondent
in the exit poll is\vadjust{\goodbreak} associated with voting patterns, then the
conclusions from their study can be dramatically altered if the
proportion of nonignorable nonrespondents happens to be large. Jim\'
enez (see below) also mentions the fact that Prado and Sans\'o appear
to ignore the fact that the sample of voters to be interviewed by exit
pollsters was not a simple random sample but rather was stratified by
gender, age category, time of day and other variables. As long as
respondents were selected randomly within stratum and as long as the
number of individuals in each stratum is proportional to the number of
persons in the population in the same stratum, the Prado and Sans\'o
analysis is adequate.

\subsection{Mart\'in}

Mart\'in's analyses are novel in that she uses what might be termed
metadata. Metadata means different things in different contexts; in the
survey context, metadata include, for example, the time it takes each
respondent to complete the survey, the number of attempts made to
contact the participant, etc. (Groves et al., \citeyear{2009Groves}). 
In summary, metadata arise from the process of conducting the election
rather than from the election itself. Mart\'in uses information on the
number of bytes of incoming and ongoing data to CNE servers, start and
close time of connections between voting centers and CNE servers, and
number of data packets in the incoming and outgoing transmissions.

Mart\'in finds unexplainable differences in the volume of information
transmitted (both outgoing and incoming) by what she calls High Traffic
Centers and Cellular Centers when compared with the Low Traffic
Centers. From a technological point of view and given election
normative, transmitted data behavior should not differ across centers.
Further, Mart\'in finds that there is a statistically significant
association between the number of votes cast in a center and the size
of the packets that were transmitted from the center to the CNE
servers; this is unexpected under the election normative that requires
that centers transmit only a total count to CNE.

These findings prompt Mart\'in to suggest that CNE servers, voting
machines or both might have been programmed to process votes in
different types of voting centers differently. Other explanations for
the differential behaviors are possible as well, and in the absence of
information about the association between transmission volume and type
of voting center, it is not possible to conclude that tampering took
place.\vadjust{\goodbreak} Mart\'in's contribution, however, is valuable in that it
highlights the use of transmission metadata and proposes approaches to
explore those data. Electronic voting systems are becoming the norm
worldwide, and therefore, forensic methods that make use of
transmission metadata might become the standard hot auditing approach.
The limitations of Mart\'in's approach can serve to inform future
protocols, so that more conclusive (perhaps even causal) conclusions
can be reached in election auditing.

\subsection{Hausmann and Rigob\'on}

Hausman and Rigob\'on use the same exit poll data that were analyzed in
the Prado and Sans\'o manu\-script. In addition, they also use the
(known) proportion of signatures in favor of holding the referendum, by
voting center. Hausmann and Rigob\'on reason as follows: both the
proportion of signatures in a voting center and the proportion of
reported YES votes collected in the exit poll at the center are
independent and noisy measurements of the vote intention of voters in
the center. Voters in a center who had earlier signed the petition for
a referendum are expected to have cast a YES vote, but for many
reasons, it is also expected that the number of actual YES votes in a
center will not be identical to the number of voters who signed the
petition. Similarly, if the exit polls are reasonably well conducted,
one would expect that in those centers where the voters cast a high
percentage of YES votes, the survey numbers would also indicate a large
proportion of YES votes. Because the noise in the two estimates of
intention of vote are due to different factors, it is also reasonable
to think that the errors in the two measurements will be uncorrelated.

The authors found that the correlation between the estimated residuals
from models where the observed number of YES votes were regressed on
either the proportion of signatures or on the predicted proportions of
YES in the exit polls were highly correlated, at least in voting
centers where by all indications, the YES should have defeated the NO
votes. By using a latent variable approach to explain this apparent
correlation, they conclude that the evidence leads to rejection of the
null hypothesis of no electronic fraud. The statistical evidence,
coupled with several other observations, lead these authors to conclude
that voting machines in about 70\% of the voting centers were
manipulated to produce official counts that did not reflect voters' intent.

In addition to the analyses described above, Hausmann and Rigob\'on\vadjust{\goodbreak}
also address the issue of selection of voting centers where a hot audit
was conducted by the government. Each voting machine provided voters
with a paper confirmation of their vote. The paper ballots were to be
put in a sealed ballot box that could later be used to audit the
accuracy of the tallies by the voting machines. An audit conducted on
the same day in which the voting took place selected what was supposed
to be a random sample of 1\% of voting machines. For these, the machine
tallys were to be compared with the paper ballots in the ballot box.
This audit was conducted in a less than satisfactory way (see the
Carter Center report of 2004). Hausmann and Rigob\'on argue that the
selection of voting machines for this audit was far from random and
indeed suggest that the CNE selected machines only from those centers
in which the CNE knew no electronic fraud had been committed.

The Hausmann and Rigob\'on conclusions that fraud did indeed occur are
somewhat of a stretch given the evidence. While it is true that the
null hypotheses of no departures from what would be expected if the
election results reflected voters' intentions is rejected, there may be
many other explanations for what was observed. Undoubtedly, the results
from these analyses are persuasive, in particular when coupled with
other facts such as the refusal of the CNE to share its random number
generator for selecting machines to be audited. In a briefing paper
published by the Center for Economic and Policy Research,
Weisbrot, Rosnick and Tucker (\citeyear{2004Weisbrot}) 
state that not only the random number generator but the source code and
other relevant material were shared by the CNE with a group of
international observers.

\subsection{Jim\'enez}

The manuscript by Jim\'enez serves both as a valuable contribution to
this issue and also as a discussion of several of the other manuscripts
we include here. Jim\'enez uses only the actual votes that were cast in
each voting machine in voting centers with two or more machines. In the
absence of a full manual count of the paper ballots, Jim\'enez proposes
that the most reliable approach consists in testing a sequence of
hypotheses that account for scenarios where irregularities were present
but are explainable by causes other than deliberate fraud.

Jim\'enez proposes to base all inference on the sampling distribution
of test statistics which can be derived by permutation of the voting
cards of each voter in each voting center. First, he assumes that the
joint\vadjust{\goodbreak} conditional distribution of outcomes per voting machine given the
total vote count in each center is a multivariate hypergeometric
distribution. If we observe $\nu$ votes in a machine, where $y$ and $n$
correspond to YES and NO votes, and where $\nu- y - n$ correspond to
OUT votes (where OUT denotes votes that are not valid for different
reasons), any such vector of size $\nu$ can be viewed as a permutation
of a vector with the same counts, but where votes are shuffled across
voters in the machine. That is, if voters are randomly assigned to
machines, then given $(\nu, y, n)$ any permutation of vote cards has
the same probability of occurring.

Jim\'enez formulates the null hypothesis of a fair referendum, where
the votes per machine correspond to a random draw from the multivariate
hypergeometric distribution indexed by $(\nu, y, n)$. A main point in
his discussion is that rejection of the null hypothesis does not imply
that the referendum was unfair; other alternatives are also possible.
He proceeds by formulating and testing several alternative hypotheses
and finally concludes that indeed, the departures from the null model
that were observed in the 2004 RR cannot be explained by innocent
mistakes or random chance.

The final conclusion from Jim\'enez's analysis is, as in the earlier
contributions, that the irregularities in the referendum introduced a
bias in favor of the winning position, and that the bias was large
enough to have resulted in the incorrect result with high probability.
Even though Jim\'enez agrees at least in part with the other
contributors, he is critical of several of the approaches that they
used to arrive at their results. One such criticism is that no one made
use of the ``full information'' that was available from the official
reports. By ``full information'' Jim\'enez refers to the nonvalid
votes and abstentions reported at the lowest electoral unit level.
While his point that all available information ought to be used in
forensic analyses is well taken, it seems that in the case of the 2004
Venezuelan RR this additional piece of data did not lead to results
that contradicted those by the other authors in this issue.

\section{Discussion}

There are many ways in which an election can be rigged, so that the
outcome does not in the end reflect voters' intentions. The increasing
popularity of electronic voting systems has allowed for the possibility
of subtle tampering that can be difficult to identify except through a
complete audit. In some instances, even completing an audit can be a
challenge\vadjust{\goodbreak} if the electronic voting machines do not produce a paper
confirmation of vote that can be saved for a manual count if the audit
becomes necessary.\looseness=1

There has been quite a lot of discussion in the recent literature on
how to define an election protocol that can reduce the opportunity for
fraud (or even innocent mistakes) and increase voter confidence in the
outcome. An example is the work by Elklit and Reynolds (\citeyear{2002Elklit}), who
propose an election assessment approach consisting of 11 different
steps (see their Table 1). For each step, Elklit and Reynolds provide
performance indicators and also variables that can be used to determine
whether the performance is adequate. Dopp (\citeyear{2009Dopp}) focuses on
post-election auditing protocols and presents a comprehensive set of
procedures to be carried out before the post-election audit begins, as
it progresses and once it has been completed. Dopp and Elklit and
Reynolds, and indeed much of the political science literature,
emphasize the procedural aspect of election assessment and auditing.
Stark (\citeyear{2008Stark}), in contrast, views the issue of designing a post-election
audit as a constrained optimization problem and provides insight on the
size of the post-election audit sample as well as a sequential testing
approach that either confirms the election outcome after a partial
audit or leads to a complete re-count.

The collection of manuscripts included in this {\it Ve\-nezuela issue}
provides good insight into some of the statistical tools that may be
useful when evaluating the integrity of an election. While the focus of
most of the work was the 2004 Presidential Recall Referendum held in
Venezuela, the major contribution of the Venezuela issue is
methodological; the manuscripts in this issue propose creative ways in
which different sources of information arising in an election can be
analyzed and interpreted to assess the election. None of the election
forensic projects described in this issue can, by themselves, provide
convincing evidence that irregularities observed in the electoral
process are due to deliberate fraud. Even Jim\'enez, who proposes the
most sophisticated (from a statistical viewpoint) methodology, is still
unable to establish that tampering occurred with certainty. The {\it
collection} of tools and conclusions, however, does paint a persuasive
picture which suggests that a battery of tests and data sources may be
more effective for election performance assessment than a single method.

Because it is so critical that the true winner is called in an
election, it would be ideal if we could design election\vadjust{\goodbreak} audit
procedures that allow causal inference. This, however, is not possible
and as a~consequence election forensics can only suggest associations.
The approach proposed by Mart\'in, however, can be amenable to a
quasi-experimental design if the properties of the transmissions
between voting machines and central servers are well understood before
the election begins; this would allow deciding if the behavior of
transmissions that are carried out during the election is surprising in
some way. Mart\'in's work shows, above all, that there are ways other
than traditional vote counting that can shed some light on the quality
of an election.

%


\begin{thebibliography}{21}

\bibitem[\protect\citeauthoryear{Balzarotti et~al.}{2008}]{2008Balzarotti}
%
\begin{bmisc}[auto:STB|2011/11/23|09:42:52]
\bauthor{\bsnm{Balzarotti},~\bfnm{D.}\binits{D.}},
\bauthor{\bsnm{Banks},~\bfnm{G.}\binits{G.}},
\bauthor{\bsnm{Cova},~\bfnm{M.}\binits{M.}},
\bauthor{\bsnm{Felmetsger},~\bfnm{V.}\binits{V.}},
\bauthor{\bsnm{Kemmerer},~\bfnm{R.}\binits{R.}},
\bauthor{\bsnm{Robertson},~\bfnm{W.}\binits{W.}},
\bauthor{\bsnm{Valeur},~\bfnm{F.}\binits{F.}} \AND
\bauthor{\bsnm{Vigna},~\bfnm{G.}\binits{G.}}
(\byear{2008}).
\bhowpublished{Are your votes really counted? Testing the
security of
real world electronic voting systems. In {\it Proceedings of the
International Symposium on Software Testing and Analysis}. Seattle,
WA. Available at
\href{http://www.cs.ucsb.edu/\textasciitilde seclab/projects/voting/issta08\_voting.pdf}
{http://www.cs.ucsb.edu/\textasciitilde seclab/projects/voting/issta08\_}
\href{http://www.cs.ucsb.edu/\textasciitilde seclab/projects/voting/issta08\_voting.pdf}{voting.pdf}}.
\bptok{imsref}%
\end{bmisc}
%
\endbibitem

\bibitem[\protect\citeauthoryear{Center}{2005a}]{2005aCarter}
%
\begin{bmisc}[auto:STB|2011/11/23|09:42:52]
\bauthor{\bsnm{Carter Center}}
(\byear{2005}a).
\bhowpublished{Observing the Venezuela Presidential Recall Referendum. Available
{at}
\href{http://www.cartercenter.org/}{http://www.}
\href{http://www.cartercenter.org/}{cartercenter.org/}}.
\bptok{imsref}%
\end{bmisc}
%
\endbibitem

\bibitem[\protect\citeauthoryear{Center}{2005b}]{2005bCarter}
%
\begin{bmisc}[auto:STB|2011/11/23|09:42:52]
\bauthor{\bsnm{Carter Center}}
(\byear{2005}b).
\bhowpublished{Final report: The Venezuela Presidential Recall Referendum. Available
{at}
 \href{http://www.cartercenter.org/}{http://www.}
 \href{http://www.cartercenter.org/}{cartercenter.org/}}.
\bptok{imsref}%
\end{bmisc}
%
\endbibitem

\bibitem[\protect\citeauthoryear{Dopp}{2009}]{2009Dopp}
%
\begin{bmisc}[auto:STB|2011/11/23|09:42:52]
\bauthor{\bsnm{Dopp},~\bfnm{K.}\binits{K.}}
\bhowpublished{(2009). Checking election outcome accuracy:
Post-election audit
sampling methods. Paper presented at the {\it81st Annual Southern Political
Science Association Conference}}.
\bptok{imsref}%
\end{bmisc}
%
\endbibitem

\bibitem[\protect\citeauthoryear{Elklit and Reynolds}{2002}]{2002Elklit}
%
\begin{bmisc}[auto:STB|2011/11/23|09:42:52]
\bauthor{\bsnm{Elklit},~\bfnm{J.}\binits{J.}} \AND
\bauthor{\bsnm{Reynolds},~\bfnm{A.}\binits{A.}}
(\byear{2002}).
\bhowpublished{{The impact of election administration on the legitimacy of emerging
democracies: A new comparative politics research agenda}.
\textit{Commonwealth and Comparative Politics}
\textbf{40}
{86--119}}.
\bptok{imsref}%
\end{bmisc}
%
\endbibitem

\bibitem[\protect\citeauthoryear{Estok, Nevitte and Cowan}{2002}]{2002Estok}
%
\begin{bmisc}[auto:STB|2011/11/23|09:42:52]
\bauthor{\bsnm{Estok},~\bfnm{M.}\binits{M.}},
\bauthor{\bsnm{Nevitte},~\bfnm{N.}\binits{N.}} \AND
\bauthor{\bsnm{Cowan},~\bfnm{G.}\binits{G.}}
(\byear{2002}).
\bhowpublished{{\textit{The Quick Count and Election Observation}. National Democratic
Institute for International Affairs, Washington}
{DC}.}
\bptok{imsref}%
\end{bmisc}
%
\endbibitem


\bibitem[\protect\citeauthoryear{European Union}{2001}]{2001European}
%
\begin{bmisc}[auto:STB|2011/11/23|09:42:52]
\bauthor{\bsnm{European Union}}
(\byear{2001}).
\bhowpublished{Preliminary statement---2 October 2001.
European Union Election Observation Mission in Bangladesh}.
\bptok{imsref}%
\end{bmisc}
%
\endbibitem

\bibitem[\protect\citeauthoryear{Groves et~al.}{2009}]{2009Groves}
%
\begin{bmisc}[auto:STB|2011/11/23|09:42:52]
\bauthor{\bsnm{Groves},~\bfnm{R.~M.}\binits{R.~M.}},
\bauthor{\bsnm{Fowler},~\bfnm{M.~J.}\binits{M.~J.}},
\bauthor{\bsnm{Couper},~\bfnm{M.~P.}\binits{M.~P.}},
\bauthor{\bsnm{Lepkowski},~\bfnm{J.~M.}\binits{J.~M.}} \AND
\bauthor{\bsnm{Singer},~\bfnm{E.}\binits{E.}}
(\byear{2009}).
\bhowpublished{\textit{Survey Methodology}.
{Wiley}, {New York}.}
\bptok{imsref}%
\end{bmisc}
%
\endbibitem

\bibitem[\protect\citeauthoryear{Hyde}{2011}]{2011Hyde}
%
\begin{bmisc}[auto:STB|2011/11/23|09:42:52]
\bauthor{\bsnm{Hyde},~\bfnm{S.~D.}\binits{S.~D.}}
(\byear{2011}).
\bhowpublished{{Catch us if you can: International election monitoring and norm
diffusion}.
\textit{American Journal of Political Science}
\textbf{55}
{356--369}}.
\bptok{imsref}%
\end{bmisc}
%
\endbibitem

\bibitem[\protect\citeauthoryear{Jacobson and Rosenfeld}{2002}]{2002Jacobson}
%
\begin{bmisc}[auto:STB|2011/11/23|09:42:52]
\bauthor{\bsnm{Jacobson},~\bfnm{A.~J.}\binits{A.~J.}} \AND
\bauthor{\bsnm{Rosenfeld},~\bfnm{M.}\binits{M.}}
(\byear{2002}).
\bhowpublished{\textit{The Longest Night}.
{Univ. California Press}, Berkeley, CA}.
\bptok{imsref}%
\end{bmisc}
%
\endbibitem

\bibitem[\protect\citeauthoryear{Lehoucq}{2003}]{2003Lehoucq}
%
\begin{bmisc}[auto:STB|2011/11/23|09:42:52]
\bauthor{\bsnm{Lehoucq},~\bfnm{F.}\binits{F.}}
(\byear{2003}).
\bhowpublished{{Electoral fraud: Causes, types and consequences}.
\textit{Annual Review of Political Science}
\textbf{6}
{233--256}}.
\bptok{imsref}%
\end{bmisc}
%
\endbibitem

\bibitem[\protect\citeauthoryear{McCarthy et~al.}{2008}]{2008Mccarthy}
%
\begin{barticle}[mr]
\bauthor{\bsnm{McCarthy},~\bfnm{John}\binits{J.}},
\bauthor{\bsnm{Stanislevic},~\bfnm{Howard}\binits{H.}},
\bauthor{\bsnm{Lindeman},~\bfnm{Mark}\binits{M.}},
\bauthor{\bsnm{Ash},~\bfnm{Arlene~S.}\binits{A.~S.}},
\bauthor{\bsnm{Addona},~\bfnm{Vittorio}\binits{V.}} \AND
\bauthor{\bsnm{Batcher},~\bfnm{Mary}\binits{M.}}
(\byear{2008}).
\btitle{Percentage-based versus statistical-power-based vote tabulation
audits}.
\bjournal{Amer. Statist.}
\bvolume{62}
\bpages{11--16}.
\bid{doi={10.1198/000313008X273779}, issn={0003-1305}, mr={2416890}}
\bptok{imsref}%
\end{barticle}
%
\endbibitem\

\bibitem[\protect\citeauthoryear{Mebane}{2007}]{2007Mebane}
%
\begin{bmisc}[auto:STB|2011/11/23|09:42:52]
\bauthor{\bsnm{Mebane},~\bfnm{W.~R.}\binits{W.~R.}}
(\byear{2007}).
\bhowpublished{{Election forensics: Statistical interventions in election
controversies. Paper prepared for presentation at the 2007}
\textit{Annual Meeting of the American Political Science Association}.
Chicago, Aug.
{30--Sep. 2}}.
\bptok{imsref}%
\end{bmisc}
%
\endbibitem

\bibitem[\protect\citeauthoryear{Norden et~al.}{2007}]{2007Norden}
%
\begin{bmisc}[auto:STB|2011/11/23|09:42:52]
\bauthor{\bsnm{Norden},~\bfnm{L.}\binits{L.}},
\bauthor{\bsnm{Burstein},~\bfnm{A.}\binits{A.}},
\bauthor{\bsnm{Joseph},~\bfnm{L.~H.}\binits{L.~H.}} \AND
\bauthor{\bsnm{Chen},~\bfnm{M.}\binits{M.}}
(\byear{2007}).
\bhowpublished{{Post-election audits: Restoring trust in
elections}. Brennan Center for Justice. August 1, 2007}.
\bptok{imsref}%
\end{bmisc}
%
\endbibitem

\bibitem[\protect\citeauthoryear{Saltman}{2006}]{2006Saltman}
%
\begin{bmisc}[auto:STB|2011/11/23|09:42:52]
\bauthor{\bsnm{Saltman},~\bfnm{R.~G.}\binits{R.~G.}}
(\byear{2006}).
\bhowpublished{\textit{The History and Politics of Voting Technology: In Quest of Integrity
and Public Confidence}. Palgrave
{McMillan}, {New York}}.
\bptok{imsref}%
\end{bmisc}
%
\endbibitem

\bibitem[\protect\citeauthoryear{Stark}{2008}]{2008Stark}
%
\begin{barticle}[mr]
\bauthor{\bsnm{Stark},~\bfnm{Philip~B.}\binits{P.~B.}}
(\byear{2008}).
\btitle{Conservative statistical post-election audits}.
\bjournal{Ann. Appl. Stat.}
\bvolume{2}
\bpages{550--581}.
\bid{doi={10.1214/08-AOAS161}, issn={1932-6157}, mr={2524346}}
\bptok{imsref}%
\end{barticle}
%
\endbibitem

\bibitem[\protect\citeauthoryear{Stark}{2010}]{2010Stark}
%
\begin{bmisc}[auto:STB|2011/11/23|09:42:52]
\bauthor{\bsnm{Stark},~\bfnm{P.~B.}\binits{P.~B.}}
(\byear{2010}).
\bhowpublished{{Risk-limiting vote tabulation audits: The importance of cluster size}.
\textit{Chance}
\textbf{23}
{9--12}}.
\bptok{imsref}%
\end{bmisc}
%
\endbibitem

\bibitem[\protect\citeauthoryear{Taylor}{2005}]{2005Taylor}
%
\begin{bmisc}[auto:STB|2011/11/23|09:42:52]
\bauthor{\bsnm{Taylor},~\bfnm{J.}\binits{J.}}
(\byear{2005}).
\bhowpublished{{Too many ties? An Empirical Analysis of the Venezuelan Recall
Referendum. Available}
{at}
 \href{http://esdata.info/pdf/Taylor-Ties.pdf}{http://}
 \href{http://esdata.info/pdf/Taylor-Ties.pdf}{esdata.info/pdf/Taylor-Ties.pdf}}.
\bptok{imsref}%
\end{bmisc}
%
\endbibitem

\bibitem[\protect\citeauthoryear{Weisbrot, Rosnick and Tucker}{2004}]{2004Weisbrot}
%
\begin{bmisc}[auto:STB|2011/11/23|09:42:52]
\bauthor{\bsnm{Weisbrot},~\bfnm{M.}\binits{M.}},
\bauthor{\bsnm{Rosnick},~\bfnm{D.}\binits{D.}} \AND
\bauthor{\bsnm{Tucker},~\bfnm{T.}\binits{T.}}
(\byear{2004}).
\bhowpublished{Black swans, conspiracy theories and the
Quixotic search
for fraud: A look at Hausmann's and Rigob\'on's analysis of Venezuela's
referendum vote. Briefing Paper, Center for Economic and Policy
Research,
Washington, DC}.
\bptok{imsref}%
\end{bmisc}
%
\endbibitem



\bibitem[\protect\citeauthoryear{Zetter}{2008}]{2008Zetter}
%
\begin{bmisc}[auto:STB|2011/11/23|09:42:52]
\bauthor{\bsnm{Zetter},~\bfnm{K.}\binits{K.}}
(\byear{2008}).
\bhowpublished{Election problems around the country. Available at
\href{http://www.wired.com/threatlevel/2008/11/election-prob-1/}{http://www.wired.com/threatlevel/2008/11/}
\href{http://www.wired.com/threatlevel/2008/11/election-prob-1/}{election-prob-1/},
Nov. 4, 2008}.
\bptok{imsref}%
\end{bmisc}
%
\endbibitem\vspace*{-2pt}

\end{thebibliography}
\end{document}